\documentclass[aps]{revtex4}
\usepackage{graphicx}
\usepackage{natbib}
\usepackage{latexsym}

\newcommand{\lsim}{\,\lower2truept\hbox{${<\atop\hbox{\raise4truept\hbox{$
\sim$}}}$}\,}
\newcommand{\gsim}{\,\lower2truept\hbox{${>\atop\hbox{\raise4truept\hbox{$
\sim$}}}$}\,}

\begin{document}

\title{Constraining the dark energy dynamics with the cosmic microwave 
background bispectrum}
\author{Fabio Giovi$^{1}$, Carlo Baccigalupi$^{1,2}$, Francesca 
Perrotta$^{1,2}$}
\address{$^{1}$SISSA/ISAS, Via Beirut 4, 34014 Trieste, Italy\\
$^{2}$LBNL, 1 Cyclotron Road, 94720 Berkeley, CA, USA}

\begin{abstract} 
We consider the influence of the dark energy dynamics at the onset of cosmic 
acceleration on the Cosmic Microwave Background (CMB) bispectrum, through 
the weak lensing effect induced by structure formation. We study 
the line of sight behavior of the contribution to the bispectrum signal at a 
given angular multipole $l$: we show that it is non-zero in a narrow 
interval centered at a redshift $z$ satisfying the relation 
$l/r(z)\simeq k_{NL}(z)$, where the wavenumber corresponds to the scale 
entering the non-linear phase, and $r$ is the cosmological comoving distance. 
The relevant redshift interval is in the range $0.1\lsim z\lsim 2$ for 
multipoles $1000\gsim l \gsim 100$; the signal amplitude, reflecting
the perturbation dynamics, is a function of the cosmological expansion
rate at those epochs, probing the dark energy equation of state
redshift dependence independently on its present value. 

We provide a worked example by considering tracking inverse power law and 
SUGRA Quintessence scenarios, having sensibly different redshift dynamics 
and respecting all the present observational constraints.
For scenarios having the same present equation of state, we find that the 
effect described above induces a projection feature which makes the bispectra 
shifted by several tens of multipoles, about 10 times more than the 
corresponding effect on the ordinary CMB angular power spectrum. 
\end{abstract}

\maketitle

\section{Introduction}
\label{i}

The recent cosmological observations indicate that 
the Universe is nearly geometrically flat, filled with structures which 
grew out of a primordial linear spectrum of nearly Gaussian and scale 
invariant perturbations; the scalar contribution, i.e. from the perturbations 
in the energy density, appears to dominate over the tensor one from 
cosmological gravitational waves. About $5\%$ of the critical energy
density is made of baryons, while the remaining dark part is supposed
to interact at most weakly with the baryons themselves, since we
observe it only indirectly, i.e. through its gravitational effects. 
The dark component appears to be 30\% pressureless, like in Cold Dark Matter 
(CDM) scenarios, dominating the gravitational potentials perturbations which 
host visible structures like galaxies or clusters. 
The remaining $70\%$ should be in some sort of vacuum energy, with negative 
pressure acting as a repulsive gravity, and responsible for a late time cosmic 
acceleration era. 

Despite of the remarkable convergence of the most important cosmological 
parameters, the picture is clearly far from being satisfactory: 
without a better insight into the nature of the dark cosmological
component, we cannot claim to have a satisfactory physical
understanding of cosmology. In particular, the evidence for a non-zero 
vacuum energy driving cosmic acceleration today (hereafter dark
energy) is a clear sign that the study of the dark cosmological component 
could have unexpected and most important consequences on cosmology and 
more generally on fundamental physics. The evidence for the dark
energy is rather robust, since it comes from two independent datasets.
The first one is represented by the observations of distant standard 
candles, namely type Ia supernovae (hereafter SNIa), directly probing the 
cosmological expansion rate at low redshifts  \cite{SNIa}; these 
observations indicate that a $70\%$ fraction of the critical energy density 
today is in the dark energy. The same result is supported, independently, 
by the observations of the Cosmic Microwave Background (CMB) anisotropies and 
Large Scale Structure (LSS). The abundance of clustered matter is
known to be at the $30\%$ level of the critical density by observing
the nearby Universe \cite{LSS}; on the other hand, the scale
subtended in the sky by the acoustic oscillations in the CMB
anisotropy angular power spectrum clearly support a flat cosmological 
geometry (see the recent results by the Wilkinson Microwave Anisotropy Probe, 
{\tt map.gsfc.nasa.gov}, hereafter WMAP, \cite{WMAP} and references therein). 
Thus, CMB and LSS together indicate again a $70\%$ of critical energy
density in the vacuum. The only way to break this evidence would be 
invoking a low value of the Hubble constant, which is excluded by the 
Hubble Space Telescope (HST) data \cite{HST}. 

In cosmology, a constant vacuum energy density is provided by the 
Cosmological Constant. While a dark matter component has well established 
support from the theories beyond the standard model of particle 
physics, a Cosmological Constant at the observed level raises two fundamental 
questions. The first is the fine-tuning: the energy scale is about 120 orders 
of magnitude less of the Planck energy density which is supposed to be the 
unification scale of all forces in the early Universe. The second question is 
the coincidence: not only the vacuum energy scale is small with respect to any 
fundamental scale, but it is such that it is comparable to the matter energy 
density today. 

Already several years ago \cite{WETTE,RP} it was realized that the evidence 
for a Cosmological Constant would have raised these crucial questions. 
To address them, the Cosmological Constant was minimally extended to a 
dynamical and inhomogeneous scalar field, evolving slowly enough to yield an 
almost constant vacuum energy out of its potential, providing cosmic 
acceleration. The latter evidence caused a renewed interest in these models 
\cite{CDF,FERRE}. In particular, it was demonstrated how the dynamics 
of this component, under suitable potential energy, can originate
attractors in the trajectory space, named tracking solutions, capable
to reach the present dark energy density starting for a in the remote 
very early Universe, thus alleviating, at least classically, the
problem of fine-tuning mentioned above \cite{LIDDL,STEIN}. The scalar
field playing the role of the dark energy was named Quintessence. This 
description allowed to investigate the relation of the dark energy
with the other cosmological components: while an explicit coupling 
with baryons is severely constrained by observations \cite{BQ}, the 
relation of the Quintessence with the gravitational \cite{EQ} and dark 
matter \cite{CQ} sectors of the fundamental Lagrangian, as well as the 
phenomenology arising from generalized kinetic energy terms \cite{KQ}
have been extensively studied. The Quintessence field has been
proposed to be related to super-symmetry and super-gravity theories 
(see \cite{SUSY} and \cite{SUGRA}, respectively, and references
therein) but we still lack a reliable candidate from fundamental physics. 

Generally speaking, despite of the large number and variety of ideas 
proposed to explain the evidence for cosmic acceleration, going from 
quantum spacetime microphysics to exotic particle physics (see 
\cite{PEEBL} and references therein), no valid solution to the
coincidence problem has been provided yet. It is likely that a
breakthrough in the understanding of the dark energy could come from the 
future high precision data on the redshift dependence cosmological 
acceleration. Indeed, almost all the candidates for the dark energy 
predict a characteristic behavior of the equation of state, $w$, as a 
function of the redshift $z$ (see \cite{CC} and references
therein). A good example of this occurrence, which we will use
extensively in this work, are the different Quintessence potentials proposed 
so far, causing relevant differences in $w(z)$ at redshift relevant
for cosmic acceleration, even if its present value is close to the 
Cosmological Constant case, $w_{0}=-1$ \cite{ERIC}. 

The recent CMB observations, combined with the LSS, HST and SNIa data, 
allow to constrain the effective dark energy equation of state, assumed to be 
constant in time, to be $w_{eff}<-0.78$ at $2\sigma$ confidence level
imposing $w>-1$ prior and $w_{eff}=-0.98\pm 0.12$ dropping the
previous prior (see \cite{WMAPCOSMPAR} and references therein). That
will be certainly improved by future observations, but the real
challenge in order to discriminate between different models is to gain 
insight into the dark energy equation of state redshift dependence. 
Together with the observations of distant SNIa, it is likely that the 
study of cosmological structure formation will be crucial for
investigating the onset of cosmic acceleration, since the redshift
intervals relevant for the two processes overlap. The best way to look
at the structure formation process is just to study the weak
gravitational lensing deflection induced by forming structures along
the line of sight of the background light (see \cite{BARTE} for
reviews). The shear induced by weak lensing on background galaxies has 
been observed by several groups, and the results agree impressively 
between different telescopes and data reduction techniques \cite{WL}; 
already, several dark energy weak lensing observables have been
proposed \cite{WLDE}. New powerful probes to map the weak lensing
shear over large sky areas are being designed to operate in the next 
decade with ground-based and space surveys (see \cite{REFRE} and
references therein). On the CMB side, the weak lensing effect on the
total intensity and polarization CMB anisotropies can be predicted on 
the general basis of cosmological perturbation theory \cite{HU}. The 
effect is generally non-Gaussian, and relevant at the arcminutes
angular scale; it is likely that the low noise, arcminute resolution 
imaging expected by the Planck satellite (see 
{\tt astro.estec.esa.nl/SA-general/Projects/Planck/}) and future CMB missions 
(see CMBpol {\tt spacescience.nasa.gov/missions/concepts.htm}) will
have a crucial importance in weak lensing studies. 

The weak lensing effect on the CMB is the subject of the present work. 
We focus on the third order statistics, where the background CMB light, 
if well described by a Gaussian process at last scattering as it appears 
from the recent CMB observations \cite{KOMAWMAP}, yields a zero contribution 
within cosmic variance. As we already mentioned, a suitable harmonic approach 
to the weak lensing effects on the CMB total intensity and polarization has 
been carried out \cite{HU}, out of a general description of the CMB 
anisotropies \cite{ISW}. Further works focused specifically on the effects on 
the CMB bispectrum \cite{KOMA}, and in particular on the benefits of having 
bispectrum data in addition to the CMB power spectrum, allowing more precise 
estimates of the main cosmological parameters, even breaking the degeneracies 
when the dark energy is included into the analysis \cite{VERDE}. In this 
work we investigate the sensitivity of the bispectrum to the redshift 
dependence of the dark energy equation of state $w(z)$. 

In Section \ref{Qdynamics} we review the basic features of the Quintessence 
dynamics in tracking trajectories, which we will use later to illustrate the 
phenomenology we study. In Section \ref{CMBbispectrum} we review the CMB third 
order statistics, described through the bispectrum, and how it is
affected through the weak gravitational lensing induced by structure 
formation. In Section \ref{lineofsight} we study the bispectrum line
of sight integral describing how it is sensitive to the cosmological 
expansion rate redshift dependence, and in Section \ref{illustration}
we provide a first quantification of the effect by considering the
leading Quintessence scenarios. Finally, in Section \ref{sac} we make
the concluding remarks.

\section{Quintessence Cosmology}
\label{Qdynamics}

The dynamics of the Quintessence scalar field $\phi$ in cosmology 
follows the Klein-Gordon equation, describing background and linear 
perturbation evolution: 
\begin{equation}
\label{KG}
\Box{\phi}+V'=0\ \ ,\ \ \delta(\Box{\phi})+V''\delta\phi=0\ ,
\end{equation}
where the $\Box$ is the Dalambertian operator in a Friedmann Robertson 
Walker (FRW) cosmology, and $\delta$ represents linear fluctuation, while 
$V(\phi)$ is the Quintessence potential and $V'=dV/d\phi$. 

\begin{figure}
\centering
\includegraphics[width=15cm,height=10cm]{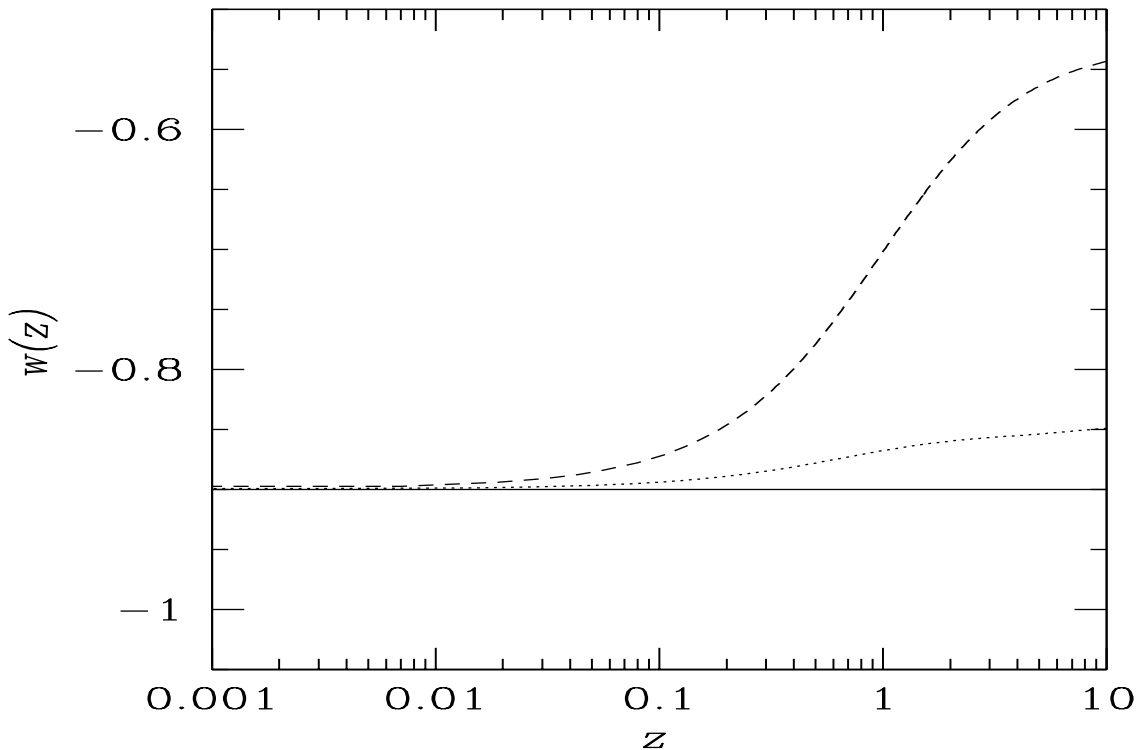}
\caption{Equation of state redshift dependence for the Quintessence models 
described in the text, for the RP model (dotted line) and the SUGRA
(dashed line). Note that both have $w_{0}$ at the present. The solid
line is a DE$_{eff}$ model with a constant equation of state.}
\label{g:w90}
\end{figure}

We consider here inverse power law (hereafter RP, first considered in 
\cite{RP}) and SUGRA \cite{SUGRA} potentials, defined respectively as 
\begin{equation}
\label{RPSUGRA}
V_{RP}={M^{4+\alpha}\over\phi^{\alpha}}\ \ ,
\ \ V_{SUGRA}={M^{4+\alpha}\over\phi^{\alpha}}\cdot \exp{\left(4\pi G
\phi^{2}\right)}\ .
\end{equation}
The exponential in the SUGRA model induces relevant changes in the 
field dynamics with respect to the RP case, as we explain now. 
As we mentioned in the Introduction, the tracking trajectories in 
the Quintessence scenarios alleviate the fine-tuning in the early Universe, 
since by means of them one is no longer forced to assume that the dark energy 
is vanishingly small with respect to radiation in the early Universe 
\cite{RP,WETTE,FERRE,STEIN,LIDDL}: the present field value $\phi_{0}$, of the 
order of the Planck mass $M_{\rm Planck}$, is reached from a wide set
of initial conditions $\phi_{i}$ and $\dot{\phi}_{i}$, with the only 
important condition that $\phi_{i}\ll M_{\rm Planck}$. Until the
latter regime is satisfied both in the RP as well as the SUGRA cases,
the tracking trajectories are well defined until the Quintessence is 
subdominating, yielding an almost constant equation of state obeying
the simple relation 
\begin{equation}
\label{track}
w={p_{DE}\over\rho_{DE}}=-{2\over 2+\alpha}\ ,
\end{equation}
where $p_{DE}=(d\phi /dt)^{2}/2-V$ and $\rho_{DE}=p_{DE}+2V$ are 
the field pressure and energy density, respectively. 
In the RP scenario, if we require the present equation of state to be in 
agreement with the current constraints \cite{WMAP}, say 
$-1\lsim w_{eff}\lsim -0.8$, the exponent is in the range 
$0\lsim\alpha\lsim 0.5$, yielding a shallow potential shape. At the present, 
the tracking regime is abandoned because the dark energy is no longer a 
subdominant component, but the shallow potential shape makes the present 
equation of state not far from the tracking one in Eq. (\ref{track}), 
differing typically at the $10\%$ level. 
The SUGRA exponential correction flattens sensibly the potential shape at 
$\phi\simeq M_{\rm Planck}$, i.e. at the {\it end} of the tracking trajectory. 
This means that a given equation of state at present is obtained for values of 
$\alpha$ sensibly larger than in the RP case, meaning that the dark energy 
dynamics and thus the cosmological expansion rate as a function of the 
redshift are generally much different in the two cases. This aspect is 
phenomenologically illustrated in Fig. \ref{g:w90}, featuring the redshift 
evolution of the equation of state in SUGRA and RP tracking trajectories 
yielding at present $w_{0}=-0.9$, from the present to the tracking regime; 
we consider a cosmological scenario which is close to the current best fit 
\cite{WMAP}, i.e. a flat Friedmann Robertson Walker (FRW) cosmology
with Hubble parameter $h=0.7$, baryon, CDM and dark energy fraction density 
$\Omega_{b}h^{2}=0.022$, $\Omega_{CDM}+\Omega_{b}=0.3$, $\Omega_{DE}=0.7$, 
respectively, three species of massless neutrinos and a scale 
invariant scalar perturbation spectrum with no gravitational waves. 
The behavior of $w(z)$ in the figure is markely different in the two 
Quintessence scenarios considered, reflecting the two values of $\alpha$ 
required to have $w_{0}$; specifically, $\alpha =0.34$ and $1.76$ in
the RP and SUGRA cases, respectively. We will also consider an
effective dark energy model with constant $w=-0.9$ (hereafter
DE$_{eff}$) obtained simply by giving that equation of state to the 
Cosmological Constant term in the FRW background cosmological
evolution. That is physically incorrect because it assumes no spatial
fluctuations in the dark energy; the latter are present indeed, and
play a relevant role even in the minimally coupled Quintessence
scenarios on super-horizon scales, as it can be seen by looking at the
large scale CMB anisotropies \cite{WELLE}; in addition, the dark
energy spatial fluctuations are present on all scales if the field
interacts with other cosmological components \cite{BQ,EQ,CQ}. The
curves in Fig. \ref{g:w90}, as well as all the perturbation spectra 
quantities in the next Section, have been numerically computed by
making use of a dark energy oriented version \cite{PERRO} of the
CMBFAST code \cite{CMBFAST}. 

The most relevant quantities for the following discussion are the distance and 
the dark energy density behavior as a function of the redshift
$z$. The first is given by 
\begin{equation} \label{e:distances}
r \left( z \right) = 
\frac{c}{H_{0}} \int_{0}^{z} \frac{dz^{\prime}}
{\sqrt{\Omega_{M0} \left( 1 + z^{\prime} \right)^{3} + \rho_{DE}(z')/
\rho_{c0}}}\ ,
\end{equation}
where we neglected the radiation contribution and $\rho_{c0}$ is the critical 
cosmological density at present; $\rho_{DE}(z)$ is the redshift dependent 
background dark energy density, given by 
\begin{equation} \label{e:omegade}
\rho_{DE} \left( z \right) = 
\rho_{DE\, 0} 
\exp \left[ 3 \int_{0}^{z} dz^{\prime} \frac{1 + w \left( z^{\prime} \right)}
{1+z^{\prime}} \right],
\end{equation}
as it can be verified by simply solving the energy conservation
equation. In the RP and SUGRA scenarios described above, $r(z)$ and 
$\rho_{DE}(z)$ have different behavior as a consequence of the
different behavior of $w(z)$ shown in Fig. 
\ref{g:w90}.

We conclude this Section by recalling briefly the main dark energy effects of 
the CMB angular power spectrum \cite{BACCI}. As $w$ gets larger than $-1$, 
the distance in Eq. (\ref{e:distances}) gets reduced. This causes a shift of 
the CMB angular power spectrum toward smaller multipoles, i.e. large angular 
scales, affecting in particular the acoustic peaks location. Moreover, for 
$w> -1$ the dark energy domination era begins earlier that in for the 
Cosmological Constant, since $\rho_{DE}$ increases with
redshifts. This boosts the gravitational potential dynamics at low 
redshift, injecting dynamics on large angular scales, where however 
the Cosmic Variance greatly limits the impact of the effect. Thus 
the projection represents the main dark energy effect on the CMB 
angular power spectrum. However, it should be noted that a limiting factor 
for the CMB power spectrum as a probe of the dark energy is the fact 
that the CMB path performs a sort of redshift average of the dark 
energy properties along their line of sight, making the CMB angular 
power spectrum mainly sensitive to the redshift average of the dark 
energy equation of state in the interval in which it is relevant, say 
$z\lsim 2$, quite low with respect to the origin of the CMB at 
$z\simeq 1000$; the latter statement could be violated if the dark 
energy is not subdominant at last scattering
\cite{BACCI}. Nevertheless, it should be noted that a great
improvement in that respect is expected by the next generation CMB 
satellites (see \cite{BALBI} for the Planck case). 

On the other hand, the injection of power in the CMB anisotropies
through the forming structures is potentially a stronger dark energy 
probe, since the relevant redshift interval is close to the onset of 
cosmic acceleration, as we pointed out already. This process is likely 
to alter the statistical properties of the CMB last scattered
radiation \cite{HU}. In particular, if the latter is Gaussian as it 
appears from the recent CMB observations \cite{KOMAWMAP}, it is
convenient to design CMB observables looking where the noise due to
the last scattered signal is lower, for example in the third order statistics. 
In the following we focus on this aspect. 

\section{Weak lensing and CMB bispectrum}
\label{CMBbispectrum}

In this Section we write down the equations describing the weak lensing effect 
induced by structure formation on the CMB bispectrum. 
We decompose the CMB anisotropy in a direction $\hat{n}$ into two terms:
\begin{equation} \label{e:dt1}
\Theta \left( \hat{n} \right) \simeq \Theta_{lss} \left( \hat{n} + 
\vec{\alpha} \right) + \Theta_{ISW} \left( \hat{n} \right)\ ;
\end{equation}
the first term represents the CMB anisotropy at decoupling ($lss$ means last 
scattering surface), which was last scattered on a direction $\hat{n}+
\vec{\alpha}$ and gravitationally lensed to our line of sight $\hat{n}$; the 
second term is the CMB anisotropy contribution from the forming structures 
along the line of sight, the Integrated Sachs-Wolfe effect (hereafter ISW 
\cite{ISW}, also known as Rees-Sciama \cite{RS}). We do not consider here the 
contributions coming from the Sunyaev-Zel'dovich effect \cite{SZ},
which yields different power at different frequencies, and the 
Ostriker-Vishniac effect \cite{OV}. To the first order in
$\vec{\alpha}$, Eq. (\ref{e:dt1}) becomes 
\begin{equation} \label{e:dt2}
\Theta \left( \hat{n} \right) \simeq \Theta_{lss} \left( \hat{n} \right) + 
\Theta_{ISW} \left( \hat{n} \right) + \vec{\nabla} \Theta_{lss} \left( 
\hat{n} \right) \cdot \vec{\alpha}\ ,
\end{equation}
and the ISW term can be expressed as 
\begin{equation} \label{e:dtrs}
\Theta_{ISW} \left( \hat{n} \right) = 2 \int dr \frac{\partial}{\partial \eta} 
\Psi \left( \eta , \hat{n} r \right)\ ,
\end{equation}
where $\eta$ is the cosmological conformal time, $r$ the comoving distance 
along the line of sight, and $\Psi$ is the cosmological gravitational 
potential, defined as the fluctuation in the metric $ds^{2}=a^{2}
\left[ \left( 1 + \frac{2 \Psi}{c^{2}} \right) dt^{2} - \left( 1 -
    \frac{2 \Psi}{c^{2}} \right) dx^{2} \right]$, if the anisotropic
stress is zero \cite{ISW,HU}. 

Note that in general Eq. (\ref{e:dtrs}) describes the 
contribution from linear and non-linear density fluctuations, with the only 
assumption that the gravitational potential fluctuations they induce are 
linear. The effect of gravitational lensing on the CMB is contained 
into the last term of Eq. (\ref{e:dt2}), indeed, $\vec{\alpha}$ is the 
gradient of the lensing potential defined as the projection of gravitational 
potential along the line of sight (see \cite{KAI,BARTE}):
\begin{equation} \label{e:lensing potential}
\phi \left( \hat{n} \right) = -2 \int^{r_{lss}}_{0} dr 
\frac{r_{lss}-r}{r_{lss}r} \Psi \left(r, \hat{n} r \right)\ .
\end{equation}
Note that the ISW and lensing effects are correlated since they 
both arise from $\Psi$; this correlation induces a non-Gaussian 
feature on the CMB pattern and leads to a non vanishing bispectrum 
signal \cite{HU,KOMA,SPERG,VERDE}. 

The gravitational potential $\Psi$ is built out of the density fluctuations 
through the Poisson equation. To describe the non-linear part of the power 
spectrum, we adopt a semi-analytic prescription \cite{MA} which evaluate the 
non-linear power out of a remapping and rescaling of the linear one, as we 
describe below; note that, as we show explicitly in the next Section, our 
results do not depend on the particular recipe we adopted for treating 
the non-linear density perturbations, which have been described in several 
ways (see for example \cite{SMITH} and references therein), but mainly 
on the perturbation dynamics in linear regime. The non-linear ($NL$)
and linear ($L$) density fluctuation power spectra are related as 
\begin{equation} \label{e:MA}
P_{NL}(k,z) = G\left[ \frac{4 \pi k^{3} P_{L}(k,z)}{f^{3/2} 
\sigma_{8}(z)^{\beta}}, z \right] 
\left( \frac{k_{L}}{k} \right)^{3} P_{L}(k_{L},z)\ ,
\end{equation}
where the linear one is evaluated at the wavelength 
\begin{equation}
\label{kl}
k_{L} = \frac{k}{\left[ 1 + 4 \pi k^{3}P_{NL}(k,z)\right]^{1/3}}\ ; 
\end{equation}
the linear power spectrum is defined as usual as 
\begin{equation} \label{e:lmps}
P_{L}(k,z)=Ak^{n}T^{2}(k,z)\ ,
\end{equation}
where $A$ is the primordial perturbation amplitude, $n$ is the
spectral index, and $T(k,z)$ is the matter transfer function. Note
that Eq. (\ref{e:MA}) with (\ref{kl}) represents an implicit equation 
which has to be solved for $P_{NL}(k,z)$ for any wavenumber $k$ at any 
epoch $z$. In the relation (\ref{e:MA}) $G$ is a function describing
the rise of the power spectrum at the onset of non-linearity;
specifically, a suitable fit is $G(x,z)=\left[ 1 + \ln \left( 1+x/2 
\right) \right] \left[ 1 + 0.02x^{4} + 1.08 \cdot 10^{-4} x^{8}/g(z) 
\right] / \left( 1 + 2.1 \cdot 10^{-5}x^{15/2} \right)$, with
$\beta=0.83$ and $f=g_{0}|w_{0}|^{1.3|w_{0}|-0.76}$ \cite{MA}. We are 
interested in the case in which the dark energy equation of state
changes with time. The latter modifies the background dynamics and 
therefore the perturbation growth rate, defined as usual as 
\begin{equation} \label{e:growth}
g(z)=g_{0}(1+z) \frac{T \left( k \rightarrow 0 ,z \right)}{T \left( k 
\rightarrow 0,0 \right)}\ . 
\end{equation}
Thus $w(z)$ enters in the linear perturbation growth rate 
as well as in the normalization $\sigma_{8}$ as a function of the redshift. 
As we mentioned in the previous Section, all the quantities above, 
i.e. $A$, $T(k,z)$ and $g(z)$, as well as $\sigma_{8}(z)$ are computed 
numerically in the Quintessence cosmologies of interest here
\cite{PERRO}; the perturbation amplitude $A$ is determined normalizing
the corresponding CMB power spectrum to the WMAP data. 

\begin{figure}
\centering
\includegraphics[width=15cm,height=10cm]{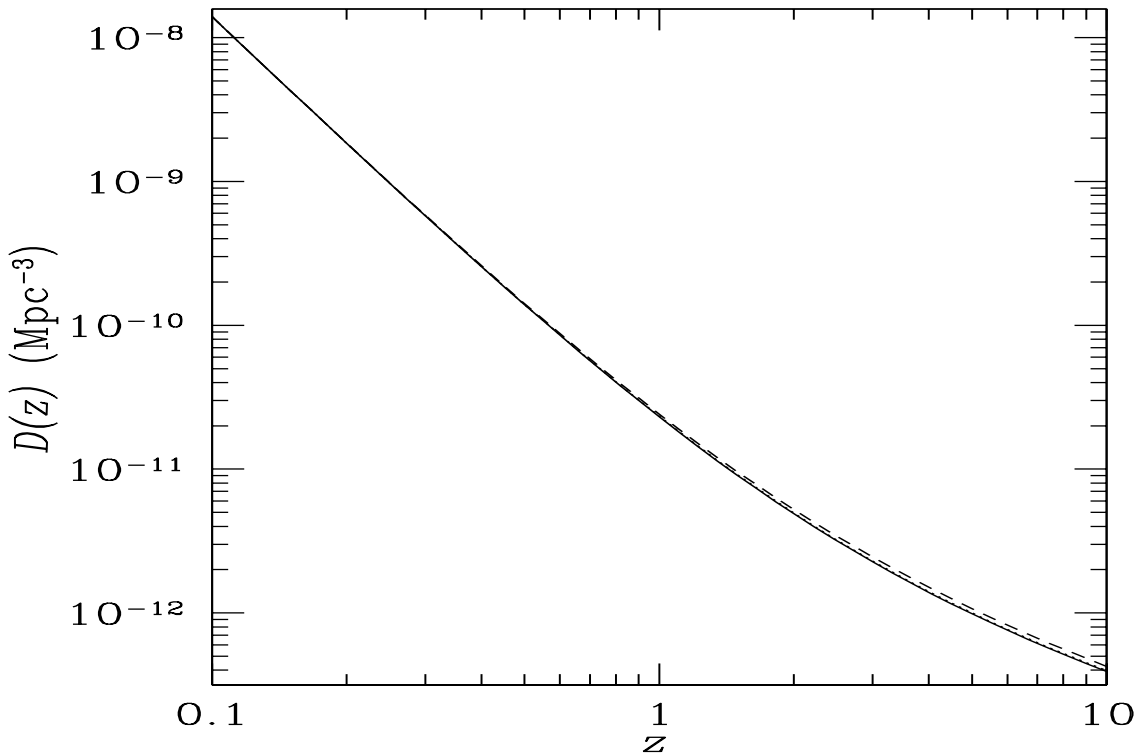}
\caption{$D$ factor as function of redshift for our three models: 
solid line is DE$_{eff}$, dotted line is RP and dashed line is SUGRA. 
The three curves, almost superposed, do not show significant
differences between the three models.}
\label{g:dfactor}
\end{figure}

The CMB bispectrum is built out of the third order statistics in the harmonic 
domain, and it is defined as 
\begin{equation}
B_{l_{1}l_{2}l_{3}}^{m_{1}m_{2}m_{3}} = a_{l_{1}m_{1}}a_{l_{2}m_{2}}a_{l_{3}
m_{3}},
\end{equation}
where $a_{lm}$ are the coefficients of the expansion of the total intensity 
CMB fluctuations into spherical harmonics; we choose to work with
dimensionless quantities when treating the CMB: in particular the
$a_{lm}$ are dimensionless corresponding to the harmonic expansion of
$\Delta T/T_{0}$ where $T_{0}$ is the present CMB thermodynamical
temperature. If the statistics of the signal is rotationally
invariant, it is convenient to work with the so called angle-averaged 
bispectrum \cite{KOMA}:
\begin{equation}
B_{l_{1}l_{2}l_{3}} = \sum_{m_{1}m_{2}m_{3}} \left(
\begin{array}{ccc}
l_{1} & l_{2} & l_{3} \\
m_{1} & m_{2} & m_{3}
\end{array}
\right) B_{l_{1}l_{2}l_{3}}^{m_{1}m_{2}m_{3}}\ ,
\end{equation}
where the parenthesis are the Wigner 3J symbols. Expanding in
spherical harmonics Eq. (\ref{e:dt2}), the expansion coefficients 
\cite{SPERG,KOMA} can be written as 
\begin{eqnarray} \label{e:almtot}
\nonumber
a_{lm} = a_{lm}^{lss} + a_{lm}^{ISW} + 
\sum_{l^{\prime} l^{\prime\prime}, m^{\prime} m^{\prime\prime}} \left( -1 
\right)^{m + m^{\prime} + m^{\prime\prime}} G^{l + l^{\prime} + l^{\prime
\prime}}_{-m + m^{\prime} + m^{\prime\prime}} \cdot \\
\cdot \frac{l^{\prime} \left( l^{\prime} + 1 \right) - l \left( l+1
  \right) + l^{\prime\prime} \left( l^{\prime\prime} + 1 \right)}{2}
\left( a_{l^{\prime}m^{\prime}}^{lss} \right)^{*} \left( 
a_{l^{\prime\prime}-m^{\prime\prime}}^{lens} \right)^{*}\ ,
\end{eqnarray}
where $a_{lm}^{lens}$ are the coefficients of the harmonic expansion of Eq. 
(\ref{e:lensing potential}); in the equation above 
$G_{m_{1}m_{2}m_{3}}^{l_{1}l_{2}l_{3}}$ is the Gaunt integral defined in 
appendix \ref{appendix}. After some algebra (see the appendix in
\cite{VERDE} for a detailed derivation), the bispectrum can be
expressed by exploiting the expression above as
\begin{eqnarray} \label{e:genbis}
\nonumber
B_{l_{1}l_{2}l_{3}} = \sqrt{\frac{\left( 2l_{1}+1 \right) \left( 2l_{2} + 1
\right) \left( 2l_{3}+1\right) }{4\pi }} \left(
\begin{array}{ccc}
l_{1} & l_{2} & l_{3} \\
0 & 0 & 0
\end{array} \right) \cdot \\
\cdot \frac{l_{1} \left( l_{1} + 1 \right) - l_{2} \left( l_{2} + 1
  \right) + l_{3} \left( l_{3} + 1 \right)}{2} C_{l_{1}}^{lss}Q
  \left(l_{3} \right) + 5P.\ ,
\end{eqnarray}
where $5P.$ indicates the permutation over $l_{1}, l_{2}$ and $l_{3}$,
$C_{l}^{lss}$ is the power spectrum of primordial CMB anisotropies, and 
\begin{equation} \label{e:ql}
Q \left( l \right) \equiv \left< \left( a_{lm}^{lens} \right)^{*} a_{lm}^{ISW} 
\right> \simeq 2 \int_{0}^{z_{lss}} dz \frac{r \left( z_{lss} \right) - r 
\left( z \right)}{r \left( z_{lss} \right)r^{3} \left( z \right)} \left[ \frac{
\partial P_{\Psi} \left( k,z \right)}{\partial z} \right]_{k=\frac{l}{r\left(z
\right)}}\ ,
\end{equation}
where $P_{\Psi} \left( k,z \right)$ is the gravitational potential
power spectrum, related to the density power spectrum by the relation 
$P_{\Psi} \left( k,z \right)=\left( \frac{3}{2} \Omega_{M0} \right)^{2}
\left( \frac{H_{0}}{ck} \right)^{4} P \left( k,z \right) \left( 1+z 
\right)^{2}$, and $z_{lss}$ is the redshift of last scattering surface; 
$Q \left( l \right)$ is the most relevant quantity here, describing how the 
forming structures along the line of sight induce the lensing on CMB photons, 
expressed as the statistical expectation of the correlation between
the ISW and lensing effects \cite{COO,SPERG,VERDE}. In particular, the 
expression above has been used to evaluate the bispectrum dependence
on the most important cosmological parameters, including an
effectively constant dark energy equation of state, and the benefits
of the bispectrum data on the estimation of the cosmological
parameters themselves \cite{VERDE}. In the next Section we study how
the contribution to $Q(l)$ is distributed along the line of sight, and
how it depends on the dark energy equation of state redshift dependence. 

\begin{figure}
\centering
\includegraphics[width=15cm,height=10cm]{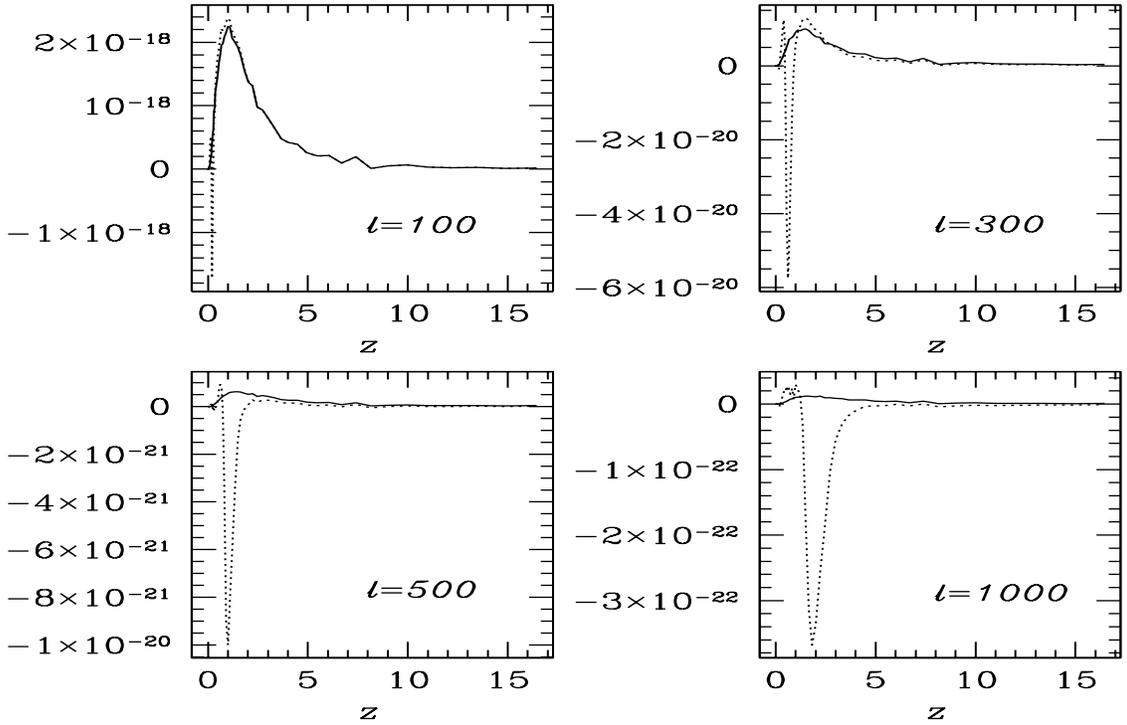}
\caption{Integrand for SUGRA model with $w_{0}=-0.9$ as function of 
redshift for different multipoles; solid line is linear regime, dotted 
line is non linear regime. The graphics are qualitative the same for 
DE$_{eff}$ and RP, the integrand is dimensionless.}
\label{g:panel}
\end{figure}

\begin{figure}
\centering
\includegraphics[width=15cm,height=10cm]{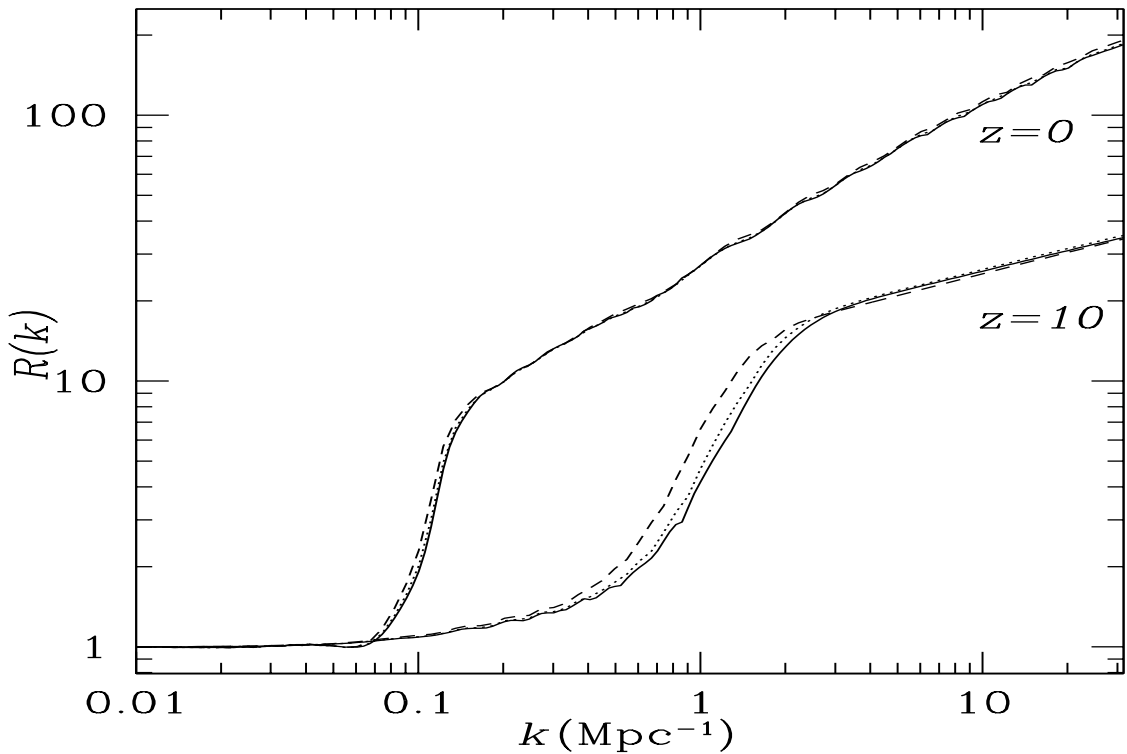}
\caption{Ratio between non linear and linear mass power spectrum for 
DE$_{eff}$ (solid line), RP (dotted line) and SUGRA (dashed line) with 
$w_{0}=-0.9$ as function of wavenumber $k$ for two redshift.}
\label{g:ratio}
\end{figure}

\section{The line of sight distribution of the bispectrum signal}
\label{lineofsight}

The integrand in Eq. (\ref{e:ql}) is built out of a geometrical factor
$D$ made of distances and a derivative factor $F$ based on the
cosmological fluctuations, this two factors are defined as 
\begin{equation}
\label{DF}
D\left( z \right) = \frac{r \left( z_{lss} \right) - r \left( z
  \right)}{r \left( z_{lss} \right)r^{3} \left( z \right)}
\ ,\  F\left( l,z \right) = \left[ \frac{\partial P_{\Psi} \left( k,z
  \right)}{\partial z} \right]_{k=\frac{l}{r\left(z\right)}}\ .
\end{equation}
In this Section we describe and study their relevance for probing the dark 
energy dynamics. 

The factor $D$ is not significantly sensitive to the dark energy redshift 
dynamics. The reason is that the dependence of $w$ on $z$ gets 
redshift averaged through the distance integral (\ref{e:distances}); the 
latter gets the dominant contribution from the lowest redshifts, say 
$z\lsim 0.1$, where all cosmologies converge to the same behavior. 
Thus, the behavior of $w$ at higher redshifts yields a negligible 
contribution to the distances, even if the differences are relevant as
in Fig. \ref{g:w90}. In Fig. \ref{g:dfactor} we plot $D$ for the three 
cosmologies defined in Section \ref{Qdynamics}, showing explicitly the 
poor sensitivity of $D$ to the $w(z)$ behavior. 

The most important contribution in the integral (\ref{e:ql}) comes from the 
gravitational potential fluctuation derivative term $F$; we now study
its behavior along the line of sight. For a given multipole $l$, and
coming from high redshift $z$, $dP_{\Psi}/dz$ is probed from large to
small scales $\lambda(z)=2\pi r(z)/l$. Let us study the asymptotic
regimes $z\rightarrow \infty$ and $z\rightarrow 0$. In the first case,
for multipoles of the order $10^{2}$, at very high $z$ the scale
$\lambda$ is outside the horizon; in this regime, if the Universe is 
matter or radiation dominated the potential is constant, so that $F$ vanishes. 
In the second case, $\lambda\rightarrow 0$, $F$ vanishes again simply because 
the power spectrum derivative is probed at infinite wavenumbers. Thus the 
relevant bispectrum signal must come from some region in between these two 
asymptotic regimes. This is evident in Fig. \ref{g:panel}, where
we plot the integrand of $Q(l)$ for different multipoles in the SUGRA 
scenario described in Section \ref{Qdynamics}, as a function of the 
redshift. As $z$ comes down from infinity, the scale $\lambda$
decreases so that eventually it meets the growing comoving scale 
corresponding to the horizon, $1/aH$; from that point $\lambda(z)$ 
becomes a sub-horizon comoving cosmological scale. In this regime, the 
gravitational potential decreases first, because of the free-streaming 
of the underlying matter density fluctuations. This feature is visible 
in Fig. \ref{g:panel} as the tail of the curve at $dP_{\Psi}/dz>0$
coming from high $z$. As $z$ gets smaller, eventually $\lambda(z)$ 
matches the scale which is entering the non-linear phase at the epoch 
$z$; at this point the derivative changes sign, $dP_{\Psi}/dz<0$, since 
the power is now increasing with time, i.e. as $z$ decreases. Thus, the 
negative peaks in Fig. \ref{g:panel} map the non-linear region of the 
spectrum. As $z$ is reduced to zero, the power vanishes approaching infinite 
wavenumbers as described above. Thus, the $F$ term yields a non-zero 
contribution to $Q(l)$ in Eq. (\ref{e:ql}) only in a redshift interval 
around the epoch corresponding roughly to the redshift $z_{bs}$
satisfying the relation 
\begin{equation}
\label{e:bingo}
{l\over r(z_{bs})}=k_{NL} \left( z_{bs} \right)\ ,
\end{equation}
where the left hand side is purely geometric, while $k_{NL}$ represents the 
scale entering the non-linear regime at $z_{bs}$. 

Now, the effect we just described turns out to be most interesting in at least 
three aspects. First, as it can be seen in Fig. \ref{g:panel}, for multipoles 
between 1000 and 100 the solution to Eq. (\ref{e:bingo}) is in a redshift 
interval extremely interesting for the onset of cosmic acceleration, say 
between $0.1$ and $2$, respectively. Second, the peak cuts out the
present epoch: this means that the bispectrum integral probes the
cosmological perturbation dynamics at the relevant redshift for cosmic 
acceleration, independently on the present. Third, for a fixed
multipole $l$, the amplitude of the signal reflects the perturbation 
dynamics at the epoch $z_{bs}$, which in turn reflects the
cosmological expansion rate at that epoch. 

In terms of the dark energy properties, the three points above mean that 
the dark energy equation of state is probed at all the relevant redshifts 
where it significantly affects the cosmological expansion rate, but 
independently on its value today; this is a very important feature
since it tends to remove the common problem which arises when trying
to constrain dark energy models with the CMB, i.e. that the relevant 
effect is made at the lowest redshifts where all the models converge
to the same behavior. We saw an example of such occurrence in the
factor $D$ above, where the distances in dark energy cosmologies are 
essentially dominated by the present dark energy equation of state
value $w_{0}$, which greatly reduces the sensitivity to the $w(z)$
behavior at higher redshifts. Indeed, that is the reason why the CMB
power spectrum is quite sensible to the effective dark energy equation of 
state $w_{eff}$, but poorly on its redshift evolution \cite{BACCI}. 
In the bispectrum, on the other hand, the signal comes from redshifts 
strictly greater than 1, cutting out the present, and this is
certainly important to discriminate between the different models
proposed to explain the evidence for cosmic
acceleration. Specifically, if $w(z)$ is different as in the models 
exposed in Section \ref{Qdynamics}, we expect to see a difference in 
the bispectrum signal, greatly enhanced with respect to the one in 
the ordinary CMB power spectrum, as we show explicitly in the next Section. 

Before concluding, we comment on the role of the description we 
adopt for the non-linear tail of the density fluctuation power spectrum. 
In Fig. \ref{g:ratio} we plot the ratio $R$ between the non-linear and linear 
power spectra according to the recipe adopted here \cite{MA}, for the dark 
energy models we consider here. The three models yield a similar behavior 
in $R$, with some difference in the scale at which $R$ rises for the onset of 
non-linearity, which in SUGRA occurs at lower wavenumbers for a fixed $z$; as 
we see in the next Section, that is not induced by the particular
recipe adopted for the non-linear mapping of the spectrum, but merely 
reflects the linear perturbation dynamics, inducing at any given epoch
a larger fluctuation amplitude in SUGRA with respect to the RP and 
DE$_{eff}$ cases, slightly shifting the non-linear rise of the
spectrum toward smaller multipoles. On the other hand, we will see
that the mechanism leading to the sensitivity of the bispectrum to the 
dark energy equation of state redshift behavior is different,
depending substantially only on the time derivative of the linear 
perturbation spectrum. 

\section{Solving for $w(z)$ with the CMB spectrum and bispectrum}
\label{illustration}

\begin{figure}
\centering
\includegraphics[width=15cm,height=10cm]{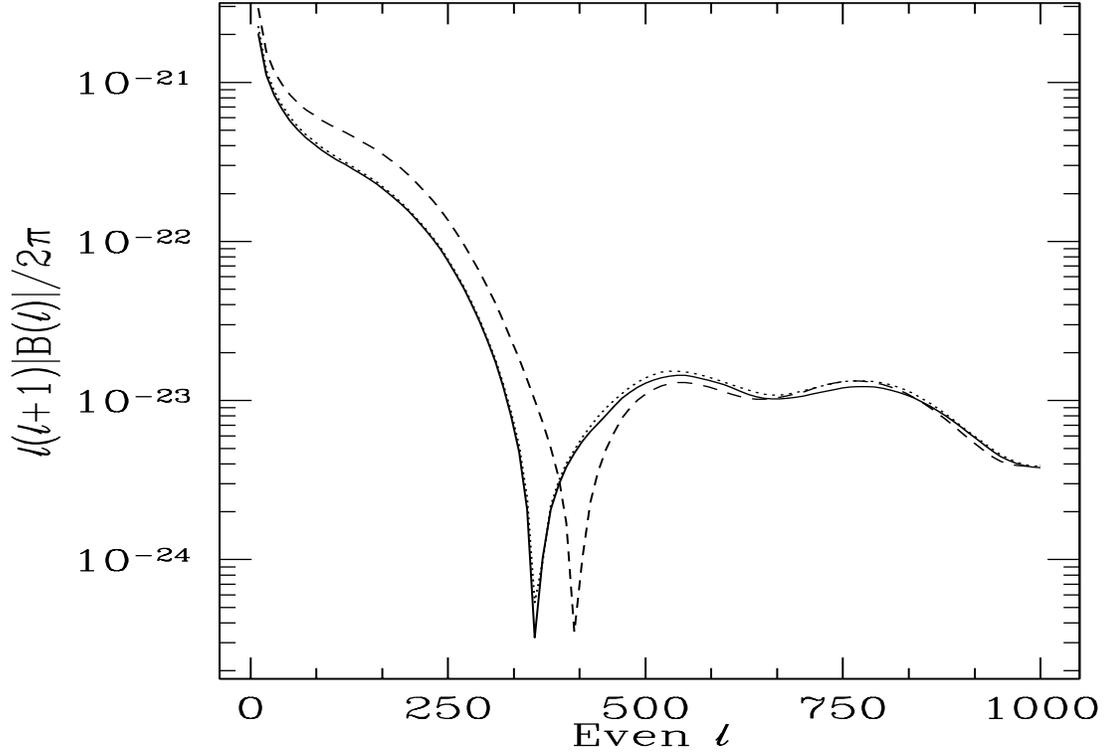}
\caption{Absolute value of dimensionless bispectrum for DE$_{eff}$ 
(solid line), RP (dotted line) and SUGRA (dashed line) with
$w_{0}=-0.9$ as function of multipole $l$.}
\label{g:bispectra}
\end{figure}

\begin{figure}
\centering
\includegraphics[width=15cm,height=10cm]{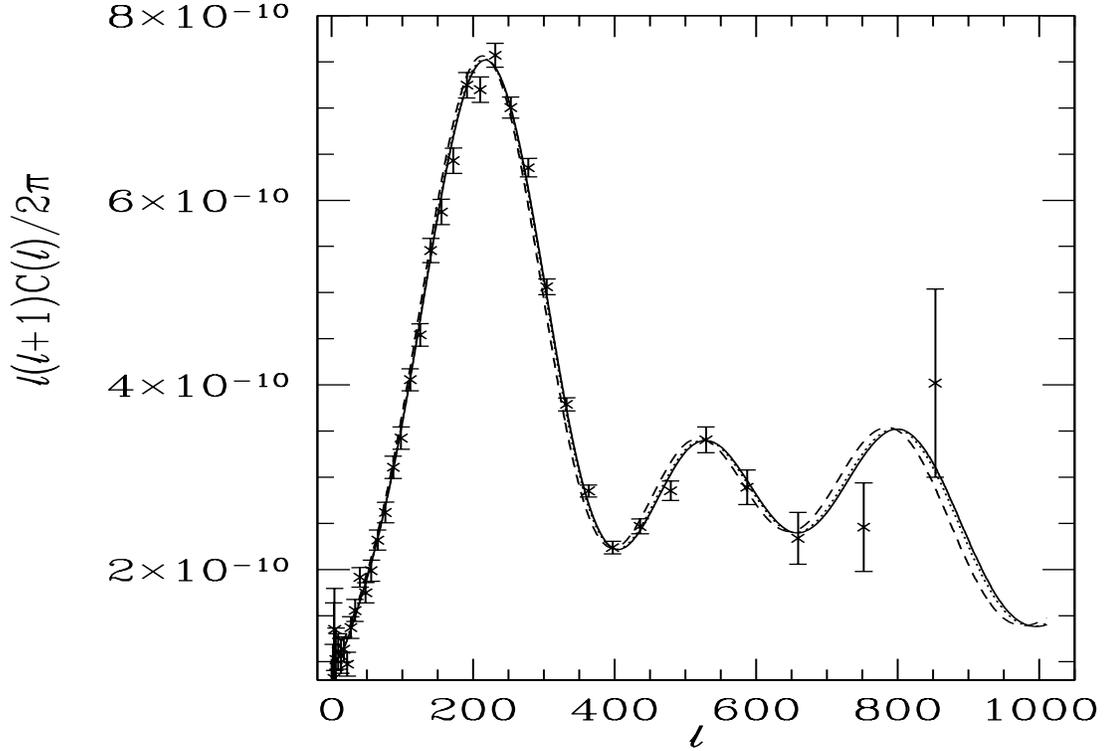}
\caption{Dimensionless power spectra for our three models as function 
of multipole $l$, notice that the degeneration between models is
greater than the bispectrum. All the models are normalized to WMAP
data. Error bars are the binned WMAP data.}
\label{g:spectra}
\end{figure}

In this Section we illustrate the phenomenology described above in the dark 
energy scenarios defined in Section \ref{Qdynamics}, by computing explicitly 
the bispectrum signal as a function of the multipole $l$. On small multipoles, 
i.e. large angles, the corresponding scales are outside the horizon, in linear 
regime, and $Q(l)>0$. On the other hand, on sufficiently large
multipoles, $Q(l)$ probes sub-horizon scales where the non-linear
regime dominates, and it has negative sign. The transition is located
at the angular scale where the contributions from the scales in linear
and non-linear regime balance in the integrand, so that $Q \left( l
\right)$ is zero. For simplicity, here we evaluate the bispectrum by 
taking equal multipoles, $l_{1}=l_{2}=l_{3}=l$. This is equivalent to probe 
the CMB pattern with equilateral triangles of different sizes corresponding 
roughly to $180/l$ degrees; that is certainly not the only choice, but it is 
enough for our purposes here. Eq. (\ref{e:genbis}) simplifies as 
\begin{equation}
B_{l}=3l \left( l+1 \right) \sqrt{\frac{\left( 2l+1 \right)^{3}}{4 \pi}} \left(
\begin{array}{ccc}
l & l & l \\
0 & 0 & 0
\end{array}
\right) C_{l}^{P} Q \left( l \right)\ .
\label{e:bispectrum}
\end{equation}
The bispectra in the three dark energy models described in Section 
\ref{Qdynamics} are plotted in Fig. \ref{g:bispectra}. In Fig. \ref{g:spectra} 
we plot the corresponding CMB angular power spectra, together with the binned 
WMAP data. As it is evident, the models are almost degenerate in the
CMB spectra, but can be resolved by looking at the bispectrum. Indeed, 
for the reason we explain in detail below, the angular location at
which the linear and non-linear bispectrum contributions balance in
the SUGRA model is shifted by several tens of multipoles toward
smaller angular scales with respect to the RP and DE$_{eff}$ cases, while in 
the ordinary CMB spectrum the shift is of the order of a few multipoles; 
this indicates that the bispectrum is almost 10 times more effective
than the ordinary CMB angular power spectrum in discriminating the 
redshift dependence of the dark energy equation of state. 

\begin{figure}
\centering
\includegraphics[width=15cm,height=10cm]{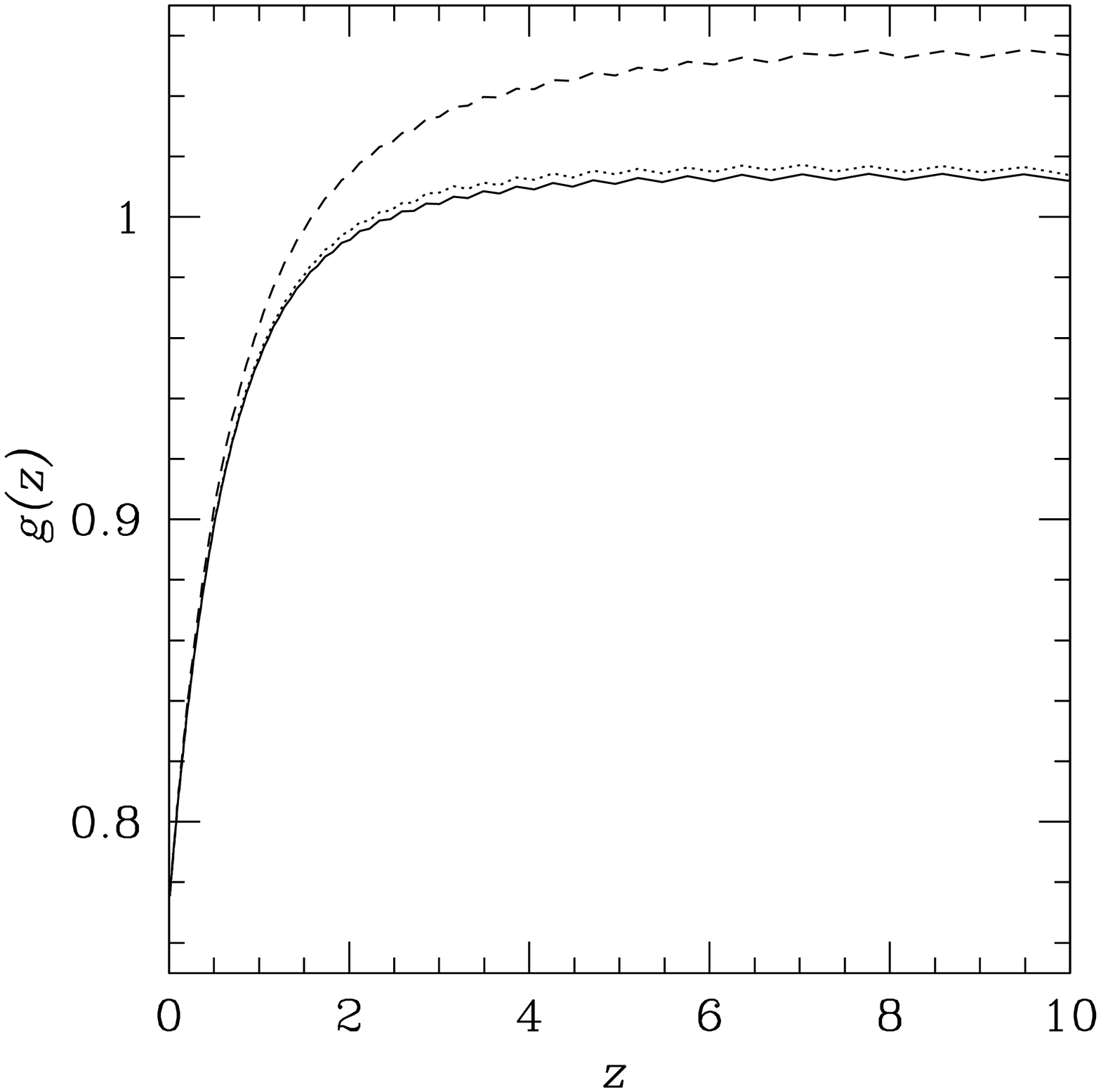}
\caption{Growth factor as function of redshift for our three models: 
DE$_{eff}$ (solid line), RP (dotted line) and SUGRA (dashed line). 
We have an appreciable difference between SUGRA and other two models
in spite of the same value of $w_{0}$.}
\label{g:growth}
\end{figure}

The three models yield equal expansion rate at the present, but as we 
pointed out in the previous Section, the bispectrum reflects the 
perturbation dynamics, and thus the cosmological expansion rate, in
the interval $0.1\lsim z\lsim 2$, independently on the present, in 
the multipole range $1000\gsim l\gsim 100$. In Fig. \ref{g:growth} we 
show the perturbation growth factor (\ref{e:growth}) in the three
models, normalized to the same value at present. As it is evident, 
the curve corresponding to the SUGRA case is sensibly larger at almost 
all epochs. Correspondingly, the change at low redshift, i.e. when the 
dark energy starts to dominate the expansion, is stronger in the 
SUGRA scenario, inducing a larger amplitude for the $dP_{\Psi}/dz >0$ 
tail in the line of sight integral (\ref{e:ql}). For the same reason, 
the bispectrum contribution coming from the gravitational bispectrum 
time derivative on non-linear scales has lower amplitude, simply
because the rise due to the onset of non-linearity has to overcome the 
gravitational potential decay, which is stronger in the SUGRA with respect 
to the RP and DE$_{eff}$ cases. 

The net effect is that, for a given multipole $l$, the larger is the
value of $w$ at the relevant epoch, the larger is the contribution
from the linear regime decay of the gravitational potential. Thus, the 
scale at which the non-linear power balances the linear one is shifted 
toward larger wavenumber, i.e. larger multipoles, as in Fig. 
\ref{g:bispectra}. Specifically, the cusp is located at $l=361,362$ for 
the DE$_{eff}$ and RP models, respectively, and at $l=412$ for
SUGRA. As we already mentioned, the corresponding shift in the CMB 
angular power spectrum in Fig.\ref{g:spectra} is just a few multipoles. 
Note that even if we increase the present equation of state in the RP
case to the border of the current constraints, say at $w_{0}=-0.78$,
the cusp moves to $l=390$, still well below the SUGRA case, simply
because no RP Quintessence model, within the current constraint, yield 
a $w(z)$ behavior as the SUGRA one in Fig. \ref{g:w90}. 

We conclude that the effect we pointed out in the previous Section represents 
a new and relevant feature induced by the dark energy dynamics on the CMB, 
which is worth taking in serious consideration in future observations: it 
causes a shift of several tens of multipoles in the CMB bispectrum arising in 
Quintessence scenarios yielding at the present the same equation of state, the 
latter being well within the range allowed by the current constraints. The 
strength of this result is evident when comparing the CMB spectra and 
bispectra in the different scenarios; the projection shift in the bispectra 
is about a factor 10 stronger than the corresponding one in the angular shift 
in the ordinary CMB spectrum. 

\section{Summary and concluding remarks}
\label{sac}

The recent evidence for cosmic acceleration is one of the greatest surprises 
in modern cosmology, and imposes the introduction of a gravitationally 
repulsive force acting on cosmological scales, known as dark energy 
\cite{SNIa,WMAP}. In the wide set of ideas proposed to explain 
this occurrence (see \cite{PEEBL}), a class of models describes the
dark energy as a self-interacting scalar field, the Quintessence, 
featuring a potential which provides the vacuum energy density
required to explain the observed level of cosmic acceleration; several 
shapes for the Quintessence potential have been proposed out of 
super-symmetry/gravity breaking models \cite{SUSY,SUGRA}. These
scenarios predict characteristic time variations of the dark energy 
equation of state $w$ \cite{CC}. By getting data on the redshift
evolution of $w$ throughout the relevant interval for cosmic
acceleration, say $z\lsim 2$, we can greatly restrict the range of
models allowed to explain the dark energy and cosmic acceleration, 
getting constraints on the fundamental physics behind this process 
\cite{ERIC}. 

From this point of view, it is important to focus on the existing cosmological 
observables and build up new ones which are able to probe the cosmological 
expansion during the onset of cosmic acceleration. The obvious, cosmological 
process which occurs at similar redshift is the structure formation, that can 
be probed statistically by the weak gravitational lensing distortions
which makes on background light \cite{KAI,BARTE}. The latter effect
has been recently observed and represents a promising probe for
cosmology \cite{WL,REFRE}. In this work we focus on the weak lensing effect 
induced by structure formation on the background CMB anisotropies 
\cite{HU,ISW}. We look at the bispectrum, built out of the third order 
moment of CMB anisotropy statistics, because if the latter is Gaussian
at last scattering as it appears from recent CMB observations 
\cite{KOMAWMAP}, then the third moment is zero within cosmic variance, 
with additional power imprinted by the weak lensing of the structures
along the line of sight of CMB photons: the latter is expressed as a 
redshift integral between the last scattering and the present
involving cosmological distances and the redshift derivative of the 
cosmological gravitational potential, induced the structures themselves 
\cite{HU,SPERG,COO,KOMA,VERDE}. 

We study the redshift behavior of the bispectrum signal arising from
the correlation between weak lensing and ISW effect; that is made by
two main contributions, deriving from the linear and non-linear
regimes of the density perturbations, yielding opposite signs after
the horizon crossing: while the linear regime makes the gravitational 
potential decreasing in time, the non-linear one, which can be
described accordingly to the existing semi-analytical recipes \cite{MA,SMITH}, 
yields a growth. Moreover, the integrand gets to zero at high and low 
redshift, and the relevant contribution, accounting for both the linear and 
non-linear regimes, comes from a narrow redshift interval centered around the 
typical epochs of structure formation, coinciding with the onset of cosmic 
acceleration. As a result, the CMB bispectrum is quite sensitive to the cosmic 
expansion rate at that redshift, i.e. on the dark energy equation of
state at that time, {\it independently} on its present value. That 
represents a new and important aspect of the CMB as a dark energy
probe, going in the direction of removing the effect coming from the
dark energy evolution at the lowest redshifts, where all the models
have to converge in order to reproduce the observed amount of cosmic 
acceleration \cite{BACCI}. To illustrate this effect, we considered two 
Quintessence scenarios both having present equation of state close to $-1$, 
but markely different in the past, as a result of the different potential 
shapes: the shallow inverse power law and the steep one including an 
exponential correction. The latter scenario generally predicts an
abrupt change of the dark energy equation of state at $z \lsim 1$,
dropping from values which can be as high as $-0.2$, while in the
first one the $w$ dynamics yield variations at the $10\%$ level. As a 
result, the gravitational potential decay in linear regime gets
enhanced in the SUGRA model with respect to the RP one; correspondingly, 
the scale at which the non-linear part of the spectrum balances the linear 
one is shifted toward gets smaller, subtending smaller angles in the sky. 
This induces a projection feature making the bispectra in the two
scenarios shifted by several tens of multipoles, i.e. 10 times stronger 
than the corresponding shift in the ordinary CMB power spectrum. 

We did not address here how much this effect is robust against variations in 
the other cosmological parameters. However, since it is mainly a projection 
feature, we argue that it could be mimicked by variations in the cosmological 
parameters inducing geometric features in the CMB anisotropies, such
as the overall cosmic geometry, Hubble constant etc. On the other
hand, it is important to note that such variations would affect
similarly {\it both} the angular power spectrum {\it and} the
bispectrum. We remark that the effect we report here affects the CMB 
bispectrum much more efficiently than the CMB spectrum, and this is
made possible by three key features which are difficult to reproduce
by varying the other cosmological parameters: $(i)$ the fact that the 
bispectrum signal induced by forming structures through weak lensing
probes a narrow redshift interval which does not include the present, 
$(ii)$ the unique prediction of the dark energy cosmology, i.e. that
one of the cosmological parameter, i.e. the equation of state, is time 
varying, and $(iii)$ that different dark energy scenarios generally
predict different time variations of the equation of state. For this 
reason we think that it will be useful to take into account the effect we 
pointed out here when dealing with data provided by the next, high 
resolution and high sensitivity cosmological probes such as Planck,
CMBpol, as well as the weak lensing surveys which will be carried out
in the next decade. 

\section*{ACKNOWLEDGMENTS}
We thank Licia Verde and Eric Linder for important suggestions.
Fabio Giovi would like to thanks Pier Stefano Corasaniti for useful 
discussions, Andrea Ferrara and Ruben Salvaterra for very important 
comments while this work was in progress. Carlo Baccigalupi is
grateful to Martin White for helpful comments. 

\appendix
\section{WIGNER 3J SYMBOLS}
\label{appendix}

The Wigner 3J symbols (W3J) come out from the composition of three 
spherical harmonics 
\begin{eqnarray} \label{e:gaunt}
\nonumber
G_{m_{1}m_{2}m_{3}}^{l_{1}l_{2}l_{3}} &=& \int d^{2}\hat{n} Y_{l_{1}m_{1}}
\left( \hat{n} \right) 
Y_{l_{2}m_{2}}\left( \hat{n}\right) Y_{l_{3}m_{3}}\left( \hat{n} \right) \\
 &=& \sqrt{\frac{\left( 2l_{1}+1 \right) \left( 2l_{2}+1\right) \left( 2l_{3}+1
\right) }{4\pi }}\left( 
\begin{array}{ccc}
l_{1} & l_{2} & l_{3} \\
0 & 0 & 0
\end{array}
\right) \left( 
\begin{array}{ccc}
l_{1} & l_{2} & l_{3} \\ 
m_{1} & m_{2} & m_{3}
\end{array}
\right) ,
\end{eqnarray}
Eq. (\ref{e:gaunt}) represent the Gaunt integral. The W3J has a closed 
formula only for a few combinations of $l_{1,}$ $l_{2}$ and $l_{3}$ and 
$m_{1,}$ $m_{2}$ and $m_{3}$; for the most general case we need a recursion 
relation, to be able to compute the whole set of W3J. The recursive
relation for W3J is computed for fixed $l_{1}$, $l_{2}$, $l_{3}$,
$m_{1}$ and for the other possible values of $m_{2}$; $m_{3}$is given
by the relation $m_{3}=-\left( m_{1}+m_{2}\right)$ and if this equality is not 
verified, the W3J are identically zero. The recursive relation \cite{SCHU} is 
\begin{equation} \label{e:recrel}
C\left( m_{2}+1\right) g\left( m_{2}+1\right) +D\left( m_{2}\right) g\left( 
m_{2}\right) +C\left( m_{2}\right) g\left( m_{2}-1\right) =0\ ,
\end{equation}
where 
\begin{equation}
g\left( m_{2}\right) =\left( 
\begin{array}{ccc}
l_{1} & l_{2} & l_{3} \\ 
m_{1} & m_{2} & m_{3}
\end{array}
\right) ,
\end{equation}
are the W3J as function of $m_{2}$, and the coefficients $C(m_{2})$
and $D(m_{2})$ are defined by
\begin{eqnarray}
C\left( m_{2}\right)  &=&\sqrt{\left( l_{2}-m_{2}+1\right) \left( l_{2}+m_{2}
\right) \left( l_{3}+m_{3}+1\right) \left( l_{3}-m_{3}\right) },
\\
D\left( m_{2}\right)  &=&l_{2}\left( l_{2}+1\right) +l_{3}\left( l_{3}+1\right)
 -l_{1}\left( l_{1}+1\right) +2m_{2}m_{3}.
\end{eqnarray}
The usual choice for the starting values of W3J at the boundaries is 
$g\left( m_{2\min}\right) = g\left( m_{2\max}\right) =1$ where $m_{2\min}$ and 
$m_{2\max}$ are respectively the smaller and the largest values of
$m_{2}$, are 
\begin{eqnarray}
m_{2\min} &=&\max \left[ -l_{2},-\left( l_{3}+m_{1}\right) \right] , \\
m_{2\max} &=&\min \left[ l_{2},l_{3}-m_{1}\right] .
\end{eqnarray}
It is also possible to work only with the forward recursion, in this way it is 
sufficient to normalize the whole set of symbols with the normalization 
condition
\begin{equation}\label{e:norm}
\sum_{m_{i}=m_{2\min }}^{m_{2\max }}\left( 2l_{1}+1\right) g^{2}\left( m_{i}
\right)=1\ ,
\end{equation}
and the phase convention
\begin{equation} \label{e:phase2}
sign\left[ g\left( m_{2\max }\right) \right] =\left( -1\right)^{l_{2}-l_{3}-
m_{1}}\ .
\end{equation}
The W3J have an analytical expression in terms of factorials, when $
m_{1}=m_{2}=m_{3}=0$ and $L=\sum\limits_{i=1}^{3}l_{i}$ is even: 
\begin{equation} \label{e:000}
\left( 
\begin{array}{ccc}
l_{1} & l_{2} & l_{3} \\ 
0 & 0 & 0
\end{array}
\right) =\left( -1\right)^{L/2}\frac{\left( L/2\right) !}{\sqrt{\left( L+1
\right) !}}\prod\limits_{i=1}^{3}\frac{\sqrt{\left( L-2l_{i}\right) !}}{\left( 
L/2-l_{i}\right) !}\ .
\end{equation}
For odd $L$ the W3J is zero. Since the computation of factorial may be hard 
at high multipoles, we can use the Gosper factorial approximation:
\begin{equation}
n! \simeq \sqrt{\left( 2n+\frac{1}{3}\right) \pi }\left( \frac{n}{e}
\right)^{n}\ .
\end{equation}
in this way we can rewrite Eq. (\ref{e:000}) in a computational simpler form
\begin{equation}
\left( 
\begin{array}{ccc}
l_{1} & l_{2} & l_{3} \\ 
0 & 0 & 0
\end{array}
\right) \simeq \left( -\frac{L}{L+1}\right) ^{L/2}\frac{1}{\left( 6L+7\right)^
{1/4}}\left( \frac{3e}{\pi }\frac{3L+1}{L+1}\right)^{1/2}\prod\limits_{i=1}^{3}
\frac{\left( 6L-12l_{i}+1\right) ^{1/4}}{\left( 3L-6l_{i}+1\right) ^{1/2}}\ .
\end{equation}

\bibliographystyle{aa}

\end{document}